\documentclass[aps,reprint,twocolumn]{revtex4-2}
\usepackage{amssymb}
\usepackage{graphicx}
\usepackage{amsmath}
\usepackage{epsfig}
\usepackage[latin9]{inputenc}
\usepackage{array}
\usepackage{multirow}
\usepackage{color}
\usepackage{esint}
\usepackage{bm}
\usepackage{bbm}
\usepackage{epstopdf}
\usepackage{mathtools}
\usepackage{soul}
\usepackage[dvipsnames]{xcolor}
\usepackage{tikz}
\usepackage{tabularx}
\usepackage{float}
\usepackage[bookmarks=true,pdftex,colorlinks=true,urlcolor=blue,linkcolor=black,citecolor=blue,breaklinks=true,hypertexnames=false]{hyperref}
\usepackage{multibib} 
\usepackage{physics}
\usepackage{makecell}
\usepackage{booktabs}
\begin{document}
\title{Thouless Pumping of Large Chern Numbers in Optical Floquet Quasicrystals}

\author{Shien Wan}
\affiliation{State Key Laboratory for Mesoscopic Physics and Frontiers Science Center for Nano-optoelectronics, School of Physics, Peking University, Beijing 100871, China}

\author{Zecheng Li}
\affiliation{State Key Laboratory for Mesoscopic Physics and Frontiers Science Center for Nano-optoelectronics, School of Physics, Peking University, Beijing 100871, China}

\author{Bo Song}
\email{bsong@pku.edu.cn}
\affiliation{State Key Laboratory for Mesoscopic Physics and Frontiers Science Center for Nano-optoelectronics, School of Physics, Peking University, Beijing 100871, China}
	
\begin{abstract}
Chern numbers are central to correlated and topological phenomena, yet most topological systems are associated with Chern numbers of order unity. Here we propose a scheme to achieve large Chern numbers in an optical Floquet quasicrystal with cold atoms, which can be directly measured via Thouless pumping. We study the quasienergy spectrum of Floquet quasicrystals and characterize the emergent Chern numbers using gap labeling theorem. We further investigate the Thouless pumping in the Floquet quasicrystal at different driving frequencies and amplitudes, revealing the connection between transport features and the quasienergy spectrum. Our findings open new avenues for exploring rich topological dynamics in Floquet quasicrystals and realizing fractional Chern insulating states.
\end{abstract}
	
\maketitle
	
\section{Introduction}

Chern numbers characterize the topological geometry of quantum states in energy bands and determine their transport properties. 
Different Chern numbers can be directly probed and separated via Thouless pumping, a quantum transport process in which the charge transported per cycle is quantized by the Chern number, with experimental realizations from photonics to ultracold atoms~\cite{thouless1983quantization,kraus2012topological,lohse2016thouless,nakajima2016topological,citro2023thouless,zhu2024reversal}.
Large Chern numbers play a critical role in strongly correlated topological matter from quantum anomalous Hall states to fractional Chern insulators~\cite{zhao2020tuning,han2024large,wan2024photoinduced,wang2013quantum,moller2015fractional,wang2022hierarchy,liu2012fractional,wang2012fractional,bernevig2025fractional}. However, realizing a topological system with robust, large Chern numbers has been a challenge. 

Here we propose a scheme for realizing large Chern numbers in an optical Floquet quasicrystal and probing them by Thouless pumping.
In contrast to periodic lattices, quasicrystals lack translational symmetry, hindering a typical definition of Chern numbers, yet they can still host exotic topology and localization~\cite{roati2008anderson,yu2024observing,kraus2016quasiperiodicity,marra2020topologically,kitaev2006anyons,yoshii2021topological}. Based on a generalized Aubry-Andr\'e-Harper (AAH) model, we directly calculate quasicrystal spectra and determine Chern numbers from the topological gap labels~\cite{aubry1980analyticity,ozawa2019topological,bellissard1992gap}. When the on-site potential modulation in the AAH model becomes incommensurate, or equivalently when the magnetic flux per plaquette is irrational, the spectrum becomes fractal assigned with a Chern number by the gap labeling theorem~\cite{9471a220-e204-3bd2-aa96-ecaca03f6cd6,bellissard1992gap,madsen2013topological}. 

Meanwhile, Floquet driving offers an alternative to engineer topological matter by coupling the energy bands~\cite{shimasaki2024reversible,eckardt2017colloquium,rudner2013anomalous} and opens large spectral gaps in quasicrystals, making the detection of large Chern numbers accessible. We investigate transport features that reflect the underlying topological spectrum at different driving frequencies and strengths in the Floquet quasicrystal.
Under off-resonant driving, atoms undergo slow Thouless pumping and diffusion emerges near the suppression of pumping. Under resonant driving, larger Chern numbers give rise to faster pumping. Our findings reveal the connection between topological spectra and Thouless pumping in optical Floquet quasicrystals and pave the way for exploring novel correlated topological matter with large Chern numbers.

\section{Optical Quasicrystal and Aubry-Andr\'e-Harper model}

The quasicrystal potential can be realized by superimposing two optical lattices, a long lattice with periodicity of $d_\mathrm{l}$ and depth of $V_{\mathrm{l}}$, and a short lattice with periodicity of $d_\mathrm{s}=\alpha d_\mathrm{l}$ and depth of $V_{\mathrm{s}}$. The resulting potential reads
\begin{equation}
	V(x,\phi) = V_{\mathrm{s}} \cos^2\!\left(\frac{\pi x}{d_\mathrm{s}}\right) 
	+ V_{\mathrm{l}} \cos^2\!\left(\frac{\pi x}{d_\mathrm{l}} - \frac{\phi}{2}\right).
    \label{eq:potential}
\end{equation}
The incommensurate $\alpha$ in the AAH model corresponds to the effective magnetic flux in units of flux quantum per unit cell in the Harper-Hofstadter model~\cite{hofstadter1976energy}. The system reduces to the Rice-Mele (RM) model for commensurate $\alpha=1/2$~\cite{lohse2016thouless,nakajima2016topological}. In the regime $V_{\mathrm{s}} \gg V_{\mathrm{l}}=2\Delta$ and under the tight-binding approximation, the system can be described by the generalized AAH model~\cite{roux2008quasiperiodic,zwerger2003mott}
\begin{equation}
\begin{aligned}
\hat{H}_{\mathrm{AAH}} 
&= -J \sum_{j} \left( \hat{c}_{j+1}^\dagger \hat{c}_{j} + \text{H.c.} \right) \\
&\quad\, + \Delta\sum_{j} \cos\bigl(2\pi\alpha j - \phi \bigr)\,\hat{c}_{j}^\dagger \hat{c}_{j},
\end{aligned}
\label{eq:H}
\end{equation}
where $J$ is the tunneling matrix element, $\Delta$ is the on-site potential.
When the hopping is a spatial constant, it generally does not close spectral gaps in quasiperiodic lattices, while in the commensurate RM model it can close gaps and change topology. Subsequent Thouless Pumping requires the localized state as the initial state.
According to its self-duality, the Anderson localization happens at $\Delta = 2J$ and in the regime $\Delta \gg J$, atoms are localized at lattice sites~\cite{aubry1980analyticity,sinai1987anderson}.

\section{Thouless Pumping and Topological Spectrum}

\begin{figure}[htbp]
    \centering
	\includegraphics[width=1\linewidth]{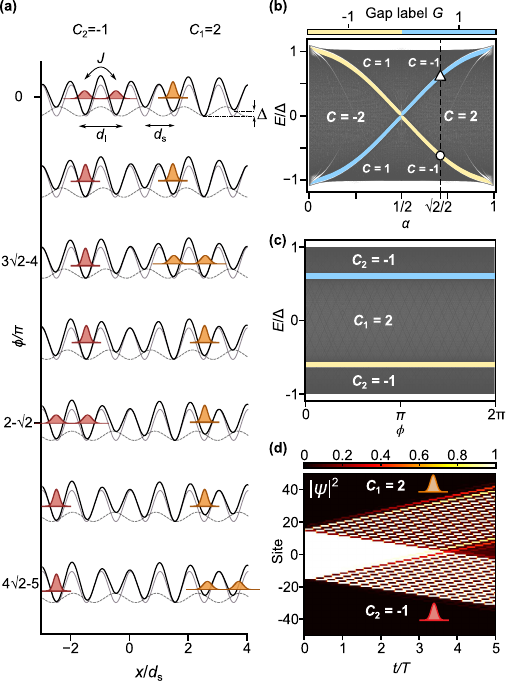}
	\caption{\textbf{Thouless pumping in an optical quasicrystal and corresponding topological quasienergy spectra.} \textbf{(a)} Optical quasicrystal (black solid line) is created by superimposing two lattices (gray dashed and solid lines) with periodicities $d_\mathrm{l}$, $d_\mathrm{s}=\alpha \,d_\mathrm{l}$ and phase difference $\phi$. Here $\alpha$ is set to $\sqrt{2}/2$. With increasing the phase $\phi$, atoms initially located at different energy belts are pumped in opposite directions with different speeds. Orange and red wavepackets represent atoms at energy belts with corresponding Chern numbers $C_1=2$ and $C_2=-1$. 
    \textbf{(b)} Spectrum of the superlattice with different $\alpha$. For $\Delta/J \gg 1$, the spectrum splits into three energy belts separated by two dominant gaps. When $\alpha$ is irrational, the potential leads to a quasicrystal. The vertical dashed line indicates $\alpha=\sqrt{2}/2$. 
    \textbf{(c)} Spectrum with three belts labeled by Chern numbers is independent of $\phi$, reflecting that the unitary period of the AAH quasicrystal tends to zero.  
    \textbf{(d)} Pumping dynamics for a uniformly filled initial state. The wavepacket splits at two distinct velocities associated with different Chern numbers.}
    \label{fig:schematic}
\end{figure}

Fig.~\ref{fig:schematic} shows atoms Thouless pumped by linearly modulating the phase $\phi$ which can be experimentally realized by an optical moving lattice with cold atoms~\cite{li2025two,tao2026optical}. Atoms initially located at different energy belts are pumped into different directions with different speeds, which is associated with corresponding Chern numbers~\cite{gottlob2025quasiperiodicity}. 

The underlying mechanism can be understood by the energy spectrum of the quasicrystal. Fig.~\ref{fig:schematic}(b) shows the spectrum of the AAH model.
For $\Delta \gg J$, two symmetric gaps of order $J$ are prominent, while other gaps are at least one order of magnitude smaller. For irrational $\alpha$, the spectrum of the AAH quasicrystal forms a Cantor set~\cite{9471a220-e204-3bd2-aa96-ecaca03f6cd6}, which makes the notion of bands no longer valid. However, the Chern numbers of each energy spectrum in the AAH quasicrystal can be determined by an extended gap labeling theorem~\cite{madsen2013topological} and are associated with spectral gaps through a generalization of the Diophantine equation $i = qm + p G$, where $i$ is the index of the magnetic subband, $p/q$ is the magnetic flux per plaquette, $m$ is integer and $G_i$ is the gap label of the $i$th gap, which is equal to the sum of the Chern numbers of all bands below it. Using the extended gap labeling theorem, the normalized integrated density of states $N(E)$ for the quasicrystal is given by taking the limit $p/q\to\alpha$,
\begin{equation}
	N(E)=\lim_{p/q\to\alpha}\frac{i}{q}
	= m + \alpha G_i,
\end{equation}
$G_i$ indicates the gap label of the spectral gap at energy $E$~\cite{bellissard1992gap,dana1985quantised,dana2014topologically,gottlob2025quasiperiodicity} and $N(E)$ is calculated by exact diagonalization. Grouping all states between two adjacent gaps into an energy ``belt'', the Chern number of the belt is defined as the difference between the gap label of the bounding gaps~\cite{thouless1982quantized},
\begin{equation}
    C_i = G_i-G_{i-1}.
    \label{eq:gap_label}
\end{equation}
Fig.~\ref{fig:schematic}(c), which shows the spectrum of the $\alpha=\sqrt{2}/2$ AAH quasicrystal. Based on the gap labeling theorem, the three belts have Chern numbers $-1,\,2,\,-1$ (See Appendix~\ref{app:numerical} for details) and are independent of $\phi$, owing to the unitary periodicity of the Hamiltonian~\cite{marra2020topologically}.

Fig.~\ref{fig:schematic}(d) shows the simulation of separated packets with different speeds as the phase is increased over time. Here we set $\Delta=0.5$ and $J=0.02$ in the unit of the recoil energy $E_r$. An initial state occupying 30 sites is chosen to approximate a uniform distribution over the phase space. The sweep rate is chosen slower than the largest gaps but faster than the remaining smaller gaps. According to the Landau-Zener tunneling, the constant flow observed in Fig.~\ref{fig:schematic}(d) throughout the pumping cycle is also consistent with the flat spectrum with respect to $\phi$ in Fig.~\ref{fig:schematic}(c) (See Appendix~\ref{app:numerical} for details).
\begin{figure}[htbp]
    \centering
    \includegraphics[width=1\linewidth]{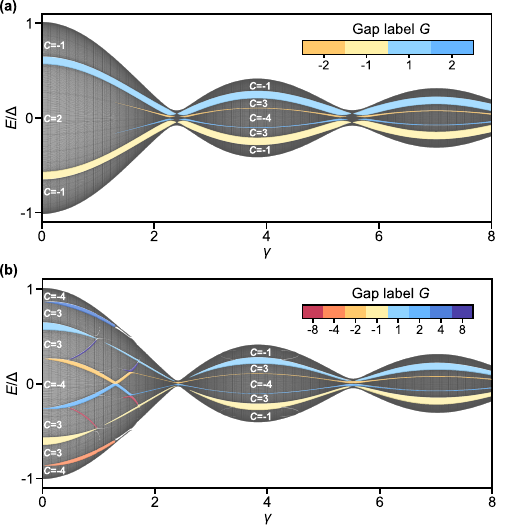}
    \caption{\textbf{Topological spectra in Floquet quasicrystals.}
    \textbf{(a)} For off-resonant modulation at $\omega\approx1.26$, three belts with Chern numbers $C=-1, 2, -1$ are separated by two dominant gaps with labels $G=-1$ and $+1$ and gap widths close to $2J$. As $\gamma$ approaches the zeros of $\mathcal{J}_0$, additional gaps with labels $G=-2$ and $+2$ emerge and the overall bandwidth is reduced to be comparable to $J$.
    \textbf{(b)} For resonant modulation at $\omega=0.3$, more Chern numbers emerge with additional arc-like gaps, as $\omega$ is comparable to the spectral bandwidth and can couple neighboring sites with energy difference $n\omega$. As $\mathcal{J}_0(\gamma)$ approaches zero, these gaps merge, while their widths scale with the $n$th-order Floquet coupling, proportional to $\mathcal{J}_n(\gamma)$. Very small gaps and the corresponding Chern numbers are not labeled for clarity.}
    \label{fig:Quasi_spectrum}
\end{figure}
This bifurcation reflects the topology of the energy spectrum, as the current equals the product of the velocity and the population, which is proportional to the Chern number. Atoms are almost reflected at the edges of the three belts and tunnel through the smaller gaps. Whenever the pumping brings neighboring on-site energies into resonance, the gap opens and the atom hops to the adjacent site. Since the belt populations remain unchanged after pumping for the uniform occupation, split atoms can continuously propagate with speeds associated with Chern numbers.

\section{Floquet Quasicrystal}

Floquet engineering can modify the hopping amplitude to induce topology and localization in both periodic and quasiperiodic lattices~\cite{lignier2007dynamical,miyake2013realizing,aidelsburger2013realization,jotzu2014experimental,bai2026exploring}. Here we implement Floquet driving, the periodic modulation of a Hamiltonian, aiming to open large gaps in optical quasicrystal and achieve measurable large Chern numbers. 
With considering Thouless pumping, the time-dependent phase modulation becomes $\phi(t) = \omega_0 t + \gamma \cos(\omega t)$ with the modulation amplitude $\gamma$.
Here we restrict the pumping dynamics to the $s$-band with $\omega \ll \Delta_{sp}$ where $\Delta_{sp}$ denotes the band gap between $s$ and $p$ bands.
Depending on the modulation frequency $\omega$, the dynamics is categorized into two regimes: an off-resonant regime ($\Delta \ll \omega \ll \Delta_{sp}$) where the Floquet expansion is valid, and a resonant regime ($J \ll \omega \lesssim \Delta$) where this expansion breaks down. Since the $s$-bandwidth is of order $\Delta$, the distinction between the off-resonant and resonant regimes is whether the drive can resonantly couple two states in the spectrum. A semi-adiabatic regime ($\omega\lesssim\omega_0$) where the phase modulation is still of the order of the linear drive is considered as the limit of the resonant regime.

\subsection{Off-resonant regime}

In the off-resonant regime, Thouless pumping under Floquet modulation shows additional pumping velocities or strong diffusion, reflecting the richer topological spectrum. The optimal frequency window for Floquet lattices satisfies $\Delta_{sp} \gg  \omega \gg J, \Delta$ ~\cite{sun2020optimal}. The condition $ \omega \ll \Delta_{sp}$ ensures that the dynamics remain within the $s$-band subspace, $ \omega \gg J,\Delta$ guarantees the convergence of the high-frequency Floquet expansion and $\omega \gg \omega_0$ allows the linear phase variation to be treated as quasi-static during the modulation cycle to second order~\cite{eckardt2017colloquium}.

Under high frequency approximation (See Appendix~\ref{app:analytics} for details), the effective Hamiltonian reads,
\begin{equation}
\begin{aligned}
\hat{H}_{0}
&= - J \sum_{j} \left( \hat{c}_{j+1}^\dagger \hat{c}_{j} + \text{H.c.} \right) \\
&\quad\,\, + \Delta \mathcal{J}_0(\gamma) \sum_{j} \cos\bigl(2\pi\alpha j - \omega_0 t \bigr)\, \hat{c}_{j}^\dagger \hat{c}_{j},
\end{aligned}
\label{eq:High-frenquency}
\end{equation}
where $\mathcal{J}_0$ is the zeroth-order Bessel function of the first kind. Eq.~\ref{eq:High-frenquency} effectively renormalizes the quasiperiodic potential strength and agrees with the quasienergy features in Fig.~\ref{fig:Quasi_spectrum}(a), obtained by diagonalizing the evolution operator $U(T')$ with $T'=2\pi/\omega$ at a typical high frequency $\omega$ as $\gamma$ varies.

The spectrum largely follows the three-belt structure of the unmodulated case, but the overall width is renormalized by $\mathcal{J}_0(\gamma)$. 
Far from the zeros of $\mathcal{J}_0$, the dynamics resembles the pumping in the static quasicrystal, showing two slow velocity branches in Fig.~\ref{fig:chern_velocity}(a). Near the zeros of $\mathcal{J}_0$, e.g. at $\gamma\simeq 2.4$ and $5.5$, the higher-order gaps of the static quasicrystal with labels $G=\pm2$ become more visible, which results in five belts with three velocity branches in Fig.~\ref{fig:chern_velocity}(b). They are $v_1=-1/(1-\alpha)$, $v_3=3/(3\alpha-2)$, and $v_4=-4/(3-4\alpha)$, with the Chern numbers of the five belts from bottom to top given by $-1$, $+3$, $-4$, $+3$, and $-1$. When $\gamma$ is tuned closer to the zeros of $\mathcal{J}_0$, such that $|\mathcal{J}_0(\gamma)|\Delta<2J$, the system can delocalize and exhibit strong diffusion, similar to ref.~\cite{shimasaki2024reversible}. See Appendix~\ref{app:numerical} for pumping dynamics at different modulation amplitudes.

\subsection{Resonant regime}

As the frequency is lowered to $\omega\lesssim\Delta$, the system enters the resonant regime, in which the high-frequency expansion no longer converges. When $\omega\gg J$, the Floquet formalism remains valid as $\omega$ remains much larger than the relevant gap sizes, and Floquet driving can open large quasienergy gaps associated with large Chern numbers, giving rise to rapid Thouless pumping. In the limiting case $\omega \sim \omega_0$, the pumping dynamics can still be treated by a semi-adiabatic approach.

For the off-resonant regime, all quasienergies
lie within the first Floquet zone $[-\omega/2,\omega/2)$, also known as the first quasienergy Brillouin zone~\cite{holthaus2016floquet}. The resonant regime involves overlapping branches from different zones. Here we use an extended Floquet-zone gauge based on the overlap of quasienergy eigenstates (See Appendix~\ref{app:numerical} for details), and obtain a clear spectrum in Fig.~\ref{fig:Quasi_spectrum}(b). Except for the $n=0$ gaps corresponding to the unmodulated quasicrystal, those opened by the $n$th-order Floquet coupling are approximately given by
$\mathcal{J}_0(\gamma)\Delta\bigl[\cos\theta-\cos(\theta+2\pi\alpha)\bigr]=n\omega$. 
Floquet couplings have cutoff frequencies and gaps thus emerge only in the resonant regime. Based on the gap-labeling theorem, the Floquet quasicrystal in the resonant regime can support gaps with large Chern numbers in Fig.~\ref{fig:Quasi_spectrum}(b). 

Notably, resonant Floquet driving plays an important role in creating an energy spectrum with large Chern numbers. Although large Chern numbers also exist in the off-resonant regime, the corresponding gap sizes are smaller. In contrast to the off-resonant driven quasicrystals or unmodulated quasicrystals in which the gaps, e.g. with labels $G=\pm2$, arise from higher-order processes and are of order $J^2/\Delta$, the gaps opened by Floquet coupling in the resonant regime are first-order effects proportional to $\mathcal{J}_n(\gamma)$ from the Fourier components of the Hamiltonian, and are of order $J$. 

\section{Topological transport}

\begin{figure}[tbp]
    \centering
    \includegraphics[width=\linewidth]{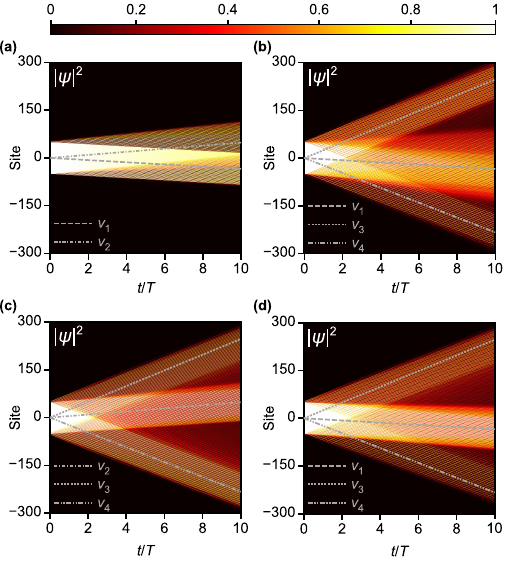}
    \caption{\textbf{Thouless pumping in Floquet quasicrystals.}
        \textbf{(a)} Pumping in the off-resonant Floquet regime at $\gamma<1$. Slow velocities $v_1$ and $v_2$ associated with $G=\pm1$ gaps are similar to the unmodulated quasicrystal.
        \textbf{(b)} Pumping in the off-resonant Floquet regime at $\gamma \approx 5$. Larger velocities $v_3$ and $v_4$ associated with $G=\pm2$ gaps emerge alongside $v_1$ from the original $G=\pm1$ gaps.
        \textbf{(c)} Pumping in the resonant Floquet regime at $\gamma<1$. Floquet driving opens more gaps and induces fast pumping with velocities $v_3$ and $v_4$. A residual $v_2$ branch appears due to incomplete semi-adiabatic separation in the dense gap structure.
        \textbf{(d)} Pumping in the resonant Floquet regime at $\gamma\approx5$. Similar to the off-resonance case, but with larger $G=\pm2$ gaps, the pumping dynamics becomes more linear.
    }
    \label{fig:chern_velocity}
\end{figure}

Fig.~\ref{fig:chern_velocity} calculated by the exact diagonalization shows the transport features of atoms at different modulation frequencies and strengths, reflecting the topological quasienergy spectrum. When the energy spectrum is uniformly occupied, the center of mass of atoms remains zero because the Chern numbers of all belts sum to zero, while each branch moves with different speeds. 
Under weakly off-resonant modulation ($\gamma<1$) in Fig.~\ref{fig:chern_velocity}(a), the pumping is similar to unmodulated quasicrystals in Fig.~\ref{fig:schematic}(d). When the modulation becomes stronger ($\gamma\approx 5$) in Fig.~\ref{fig:chern_velocity}(b), The high-velocity branches emerges from higher-order gaps of the unmodulated quasicrystal, which are enlarged by Floquet driving. The tunneling interval, the time of atoms staying in the same site per pumping cycle, are equal and independent on $\gamma$. 
Under weak resonant modulation ($\gamma<1$), the fast pumping in Fig.~\ref{fig:chern_velocity}(c) originates from larger Floquet-induced gaps in Fig.~\ref{fig:Quasi_spectrum}(b). Although the velocities are similar to the off-resonant regime, the tunneling intervals are staggered and dependent on $\gamma$. similarly, for stronger modulation ($\gamma\approx5$), the high-velocity branches in Fig.~\ref{fig:chern_velocity}(d) also originally arise from the unmodulated quasicrystal, which still enlarged by Floquet driving.

The transport velocity of each branch is determined by Chern numbers. 
For example, at $\omega=0.3$ and near $\gamma\simeq0$, the first-order Floquet driving opens gaps with labels $G=\pm2$ and $G=\pm4$, dividing the spectrum into seven belts with Chern numbers $-4$, $+3$, $+3$, $-4$, $+3$, $+3$, and $-4$ from bottom to top in Fig.~\ref{fig:Quasi_spectrum}(b). These seven belts correspond to the two velocities $v_3=3/(3\alpha-2)$ and $v_4=-4/(3-4\alpha)$ shown in Fig.~\ref{fig:chern_velocity}(c). At lower driving frequencies $\omega$, higher-order Floquet couplings can open more gaps, introducing belts with larger Chern numbers and correspondingly larger pumping velocities. From the gap labeling theorem, the velocity can be written as $1/v_{|C|}=\alpha-t_{|C|}/|C|$. Therefore, the allowed velocities are restricted to a discrete set of values. For example, for $|C|=1,2,3,4,7,10$, one obtains $t_1=1$, $t_2=1$, $t_3=2$, $t_4=3$, $t_7=5$, and $t_{10}=7$. These integers make $1/v_{|C|}+t_{|C|}/|C|$ the best approximation to $\alpha$ with fixed denominator $|C|$, which can be proved from the spectrum symmetry and the uniqueness of the decomposition $m+|C|\alpha$ for a energy belt with a given $|C|$.

For even lower driving frequencies entering the semi-adiabatic regime $\omega \lesssim \omega_0$, the pumping trajectory behaves oscillatory (See Appendix~\ref{app:numerical}). These oscillations arise from varying hopping intervals on different branches. As $\gamma$ is increased further, the system is out of the semi-adiabatic regime. Stable pumping trajectories vanish due to Landau-Zener tunneling, leading to rapid inter-belt transitions and reducing the pumping velocity. The overall probability distribution remains bounded by the gray dashed envelopes of the $v_1$ and $v_2$ branches in Fig.~\ref{fig:low_frequency}.

\begin{figure*}[!ht]
    \centering
    \includegraphics[width=\linewidth]{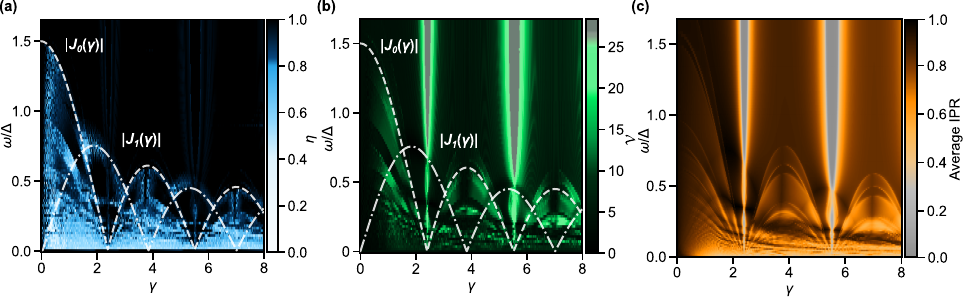}
    \caption{
        \textbf{Topological transport features in Floquet quasicrystals under different driving.}
        \textbf{(a)} In the off-resonant regime, diffusion exponent $\eta$ is generally close to 1, as the dynamics is similar to pumping in unmodulated quasicrystals or dominated by strong expansion near the zeros of $\mathcal{J}_{0}(\gamma)$. In the resonant regime, the emergence of many smaller gaps breaks the semi-adiabatic condition, leading to $\eta<1$. This sublinear region is mainly below the $|\mathcal{J}_{0}(\gamma)|$ envelope associated with the lowest-order Floquet coupling.
        \textbf{(b)} In the resonant regime, multiple fast velocities $\mathcal{V}$ reflect large Chern numbers, of which envelopes are proportional to $|\mathcal{J}_{0}(\gamma)|$ with different coefficients (dashed line). The light green stripe of fast pumping associated with the $n=\pm1$ Floquet coupling is interrupted near the zeros of $|\mathcal{J}_{1}(\gamma)|$ (dashed dotted line). At these zeros, the coupling strength vanishes, leading to closing corresponding gaps and vanishing the fast pumping.
        \textbf{(c)} Small average IPR indicates more gap opening of higher-order Floquet coupling, leading to a distribution that closely resembles that of the transport velocity $\mathcal{V}$.
    }
    \label{fig:diffusion}
\end{figure*}

We further characterize topological transport by the cloud size $\sigma(t)=\sqrt{\langle x^2\rangle-\langle x\rangle^2}$ with $x$ the position of each atom. The size growth is extracted by fits $\sigma(t)=\sqrt{\sigma^2(0)+\mathcal{V}^2 t^{2\eta}}$ with the initial cloud size $\sigma(0)$. The coefficients $\mathcal{V}$ in units of $d_\mathrm{s}/T$ and $\eta$ characterize the pumping velocity and diffusive behaviors.
$\eta=1$ indicates a linear pump and satisfies the semi-adiabatic condition. 
Fig.~\ref{fig:diffusion} shows the extracted parameters as a function of modulation frequency and strength. See Appendix~\ref{app:methods} for details of parameters.

For the off-resonant regime, e.g. high-frequency modulation beyond $\omega/\Delta \approx 1.5$, the quasienergy spectrum largely retains the three-belt structure and the pumping trajectories are almost linear with $\eta\approx 1$ and small velocities, few $d_\mathrm{s}/T$. Near the zeros of $\mathcal{J}_0$, the system exhibits strong spread with large $\mathcal{V}$ indicated in gray in Fig.~\ref{fig:diffusion}(b). For the resonant regime, i.e. under low-frequency modulation where additional gaps are opened and higher-order pumping processes emerge, richer structures develop in $\mathcal{V}$ and $\eta$. In this regime, the diffusion exponent $\eta$ deviates from unity, reflecting the breakdown of the semi-adiabatic condition due to the increasingly complex gap structure. Additional high-velocity regions beyond the zeros of $\mathcal{J}_0$, bounded by dashed curves proportional to $|\mathcal{J}_0(\gamma)|$, arise from additionally opened high-order Floquet gaps and correspond to large velocities, e.g. $v_3$ and $v_4$ on the order of $20\,d_\mathrm{s}/T$ indicated in light green in Fig.~\ref{fig:diffusion}(b).

We further calculate the average inverse participation ratio (IPR) over the Floquet quasienergy eigenstates in Fig.~\ref{fig:diffusion}(c). This is obtained by diagonalizing the evolution operator and the result is independent of $\phi$ due to the flat spectrum of the quasicrystal (See Appendix~\ref{app:numerical} for details). Interestingly, the pattern of the average IPR is similar to $\mathcal{V}$. Near boundaries of gap in the spectrum, the coupling between neighboring sites hybridizes states, leading to more extended states and smaller IPR. Smaller IPR values thus serve as an indicator of higher-order gap openings, which give rise to large pumping velocities.
	
The $|\mathcal{J}_0|$-type dependence of the average IPR can be understood from the gap-opening curves in parameter space given by Eq.~\eqref{eq:cut_off} in Appendix. These curves determine the boundaries where higher-order Floquet couplings open large-Chern-number gaps and thus give rise to stronger pumping. Different Floquet-coupling orders produce $|\mathcal{J}_0|$-like envelopes with different amplitudes, explaining the emergence of different stripes. Furthermore, the interruptions within the strong-pumping stripes, such as those dark regions observed at the zeros of $\mathcal{J}_1(\gamma)$ for the $n=\pm1$ channels in Fig.~\ref{fig:diffusion}(b) and (c), originate from the fact that the corresponding coupling strength is proportional to $\mathcal{J}_1(\gamma)$. At the zeros of $\mathcal{J}_1(\gamma)$, the coupling vanishes and no gap is opened, leading to an increase in the average IPR and the disappearance of the associated higher-order pumping channel.
	
\section{Conclusion}
We propose an optical Floquet quasicrystal scheme to realize large Chern numbers and probe them through Thouless pumping. Quasicrystals provide a novel platform for exploring large Chern numbers owe to their broad topological spectra and flat belt nature. We highlight the role of Floquet driving in opening large gaps in quasicrystals, enabling large and robust Chern numbers that are experimentally accessible. The quasienergy spectrum of the Floquet quasicrystal is calculated and unfolded by an extended Floquet-zone gauge. We determine the corresponding Chern numbers of the quasienergy spectrum using the extended gap labeling theorem. Furthermore, we investigate the transport features under different modulation frequencies and strengths. The observed Bessel-type pumping regions are attributed to resonant Floquet driving and associated Chern numbers. The Floquet quasicrystal proposed in this work paves the way for exploring new strongly correlated insulating phases beyond those realized in continuum Landau levels~\cite{moller2015fractional,wang2022hierarchy,liu2012fractional,wang2012fractional,bernevig2025fractional}.

\section{Acknowledgments} 
We thank Botao Wang, Hongzheng Zhao and Huaqing Huang for helpful discussions. This work was supported by the Quantum Science and Technology-National Science and Technology Major Project (2024ZD0301800), the Beijing Natural Science Foundation (Z240007), the National Natural Science Foundation of China (12374242), and Scientific Research Innovation Capability Support Project for Young Faculty (ZY2025014).
	
\section{Data Availability}
The data presented in this work are available on the Peking University Open Research Data Platform.

\appendix
\section{\MakeUppercase{Optical Floquet Quasicrystals}}
\label{app:analytics}

\subsection{GAAH model from perturbation theory}
\label{subsec:GAAH_pert}

The potential of a one-dimensional bichromatic optical lattice is given by
\begin{equation}
    V(x,\phi)
    =
    V_{\mathrm{s}} \cos^2\!\left(\frac{\pi x}{d_\mathrm{s}}\right)
    +
    V_{\mathrm{l}} \cos^2\!\left(\frac{\pi x}{d_\mathrm{l}} - \frac{\phi}{2}\right),
\end{equation}
where $V_{\mathrm{s}}$ and $V_{\mathrm{l}}$ are the depths of the short and long lattices and
$
\alpha = d_\mathrm{s}/d_\mathrm{l}
$
is the lattice-spacing ratio. 

In the tight-binding limit $V_{\mathrm{s}} \gg E_{\mathrm{r,s}}$, with $E_{\mathrm{r,s}}$ the recoil energy of the short lattice, the short lattice defines well localized sites at
$
x_i^{(0)} = (i-\tfrac12)d_\mathrm{s}
$,
and the $s$-band tight-binding Hamiltonian reduces to
$
-J \sum_i (\hat c_{i+1}^\dagger \hat c_i + \text{H.c.})
$ when $V_{\mathrm{l}}=0$. In the presence of the long lattice, the on-site energies are modified and the site positions are shifted, resulting in a change of the tunneling and here we consider the lowest order when $V_{\mathrm{l}}\ll V_{\mathrm{s}}$.

The leading contribution of the long lattice at $x_i^{(0)}$ reads
\begin{align}
    V_i
    &\approx
    V_{\mathrm{l}} \cos^2\left(
    \frac{\pi x_i^{(0)}}{d_\mathrm{l}} - \frac{\phi}{2}
    \right)
    \notag\\
    &=
    \frac{V_{\mathrm{l}}}{2}
    \Bigl[
    \cos\bigl(2\pi\alpha i - \phi\bigr) + 1
    \Bigr].
    \label{eq:deltaVj}
\end{align}
Note that the shift of $V_{\mathrm{s}}$ due to the small displacement of $x_j$ only happens at second order~\cite{roux2008quasiperiodic}, since $x_j^{(0)}$ is located at the short lattice minima.

The dominant modification of hopping comes from the shift of the site positions induced by $V_{\mathrm{l}}$. The nearest-neighbor hopping $J$ depends exponentially on the distance between adjacent sites~\cite{zwerger2003mott}, so the lowest-order correction to $x_i$ is
\begin{equation}
    \delta x_i
    =
    \frac{\alpha}{2\pi}\,\frac{V_{\mathrm{l}}}{V_{\mathrm{s}}}
    \sin\bigl(\pi\alpha(2i+1)-\phi\bigr)\, d_\mathrm{s}.
    \label{eq:deltaxj}
\end{equation}
The distance between neighboring sites becomes
$
x_{i+1}-x_{i}
= d_\mathrm{s} + \delta x_{i+1}-\delta x_{i}
$,
which gives the correction to hopping,
\begin{align}
    J_i
    &\simeq
    J
    \left[
    1
    - \frac{\alpha}{2}
    \sqrt{\frac{V_{\mathrm{l}}^2}{E_{\mathrm{r,s}}V_{\mathrm{s}}}}\,
    \cos\bigl[\pi\alpha(2i+1)-\phi\bigr]
    \sin(\pi\alpha)
    \right].
    \label{eq:Jj_final}
\end{align}

Using the on-site contribution~\eqref{eq:deltaVj} and the tunneling modulation~\eqref{eq:Jj_final}, the generalized Aubry--Andr\'e--Harper (GAAH) model is given as follows,
{\small
    \begin{align}
        \hat{H} = 
        &-J \sum_{i} \Biggl\{ 1 
        -\sqrt{\frac{\alpha^2V_{\mathrm{l}}^2}{4E_{\mathrm{r,s}}V_{\mathrm{s}}}}\,
        \cos\bigl[\pi\alpha(2i+1) - \phi\bigr]\sin(\pi\alpha) \Biggr\}
        \hat{c}_{i+1}^\dagger \hat{c}_{i}
        \notag\\
        & + \text{H.c.}\notag\\
        & + \frac{V_{\mathrm{l}}}{2}\sum_{i} 
        \Bigl[
        \cos\bigl(2\pi\alpha i - \phi\bigr) + 1
        \Bigr] 
        \hat{c}_{i}^\dagger \hat{c}_{i}.
        \label{eq:GAAH_final}
    \end{align}
}

After redefining the zero of energy, and introducing
\begin{equation*}
    \Delta=\frac{V_\mathrm{l}}{2},
    \qquad
    \Delta J=-\frac{\alpha J}{2}\sqrt{\frac{V_\mathrm{l}^2}{E_rV_\mathrm{s}}}\sin(\pi\alpha),
\end{equation*}
we define
\begin{equation}
    \begin{aligned}
        \Delta J_i(\phi)
        &=\Delta J\cos\bigl(\pi\alpha(2i+1)-\phi\bigr),\\
        \Delta_i(\phi)&=\Delta\cos\bigl(2\pi\alpha i-\phi\bigr).
    \end{aligned}
    \label{eq:DeltaJ_Delta}
\end{equation}
Finally, the GAAH model can be written as
\begin{equation}
    \begin{aligned}
        \hat H_{\mathrm{GAAH}}
        =&
        -\sum_i \left[J+\Delta J_i(\phi)\right]\hat c_{i+1}^\dagger \hat c_i
        +\mathrm{H.c.}\\
        &+\sum_i \Delta_i(\phi)\,\hat c_i^\dagger \hat c_i .
    \end{aligned}
    \label{eq:generalized_AAH}
\end{equation}

When the hopping modulation remains sufficiently small, the Hamiltonian can be simplified to the AAH model in the main text.

\subsection{Equivalence between AAH model and Harper-Hofstadter(HH) model}

The equivalence between the one-dimensional (1D) Thouless pumping and the two-dimensional (2D) quantum Hall effect is established by identifying the pump phase $\phi$ in the AAH model with the quasimomentum along the second dimension of the HH model. The inverse Fourier transform of $H_{\mathrm{GAAH}}$ on $\phi$ then gives a 2D lattice model in a uniform magnetic field, with both nearest-neighbor and next-nearest-neighbor hoppings
\begin{equation}
    \begin{aligned}
        \hat H_{2\mathrm D}
        ={}&
        -\sum_{m,n} J\,\hat c_{m+1,n}^\dagger \hat c_{m,n}\\
        &+\frac{\Delta}{2}\sum_{m,n} e^{+i2\pi\alpha m} \hat c_{m,n+1}^\dagger \hat c_{m,n}
        \\
        &-\frac{\Delta J}{2}\sum_{m,n} e^{+i2\pi\alpha (m+1/2)} \hat c_{m+1,n+1}^\dagger \hat c_{m,n}
        \\
        &-\frac{\Delta J}{2}\sum_{m,n} e^{-i2\pi\alpha (m+1/2)} \hat c_{m+1,n}^\dagger \hat c_{m,n+1}\\
        &+\mathrm{H.c.}
        \label{eq:GAAH_2D}
    \end{aligned}
\end{equation}

Considering the AAH model without hopping modulation, Eq.~\eqref{eq:GAAH_2D} reduces to the HH model with only nearest-neighbor hopping and a magnetic flux $\Phi=\alpha\Phi_0$ through each plaquette
\begin{equation}
    \hat H_{HH}
    =
    \sum_{m,n}\left[
    -J\,\hat c_{m+1,n}^\dagger \hat c_{m,n}
    +\frac{\Delta}{2} e^{i2\pi\alpha m} \hat c_{m,n+1}^\dagger \hat c_{m,n}
    \right]+\mathrm{H.c.}
    \label{eq:HH}
\end{equation}
With Floquet driving, the Harper-Hofstadter model becomes Floquet Harper-Hofstadter model~\cite{bai2026exploring},
\begin{equation}
    \begin{aligned}
        \hat H_{\mathrm{FHH}}(t)
        =&
        -\sum_{m,n} J\,\hat c_{m+1,n}^\dagger \hat c_{m,n}\\
        &+\frac{\Delta}{2}\sum_{m,n}
        e^{i[2\pi\alpha m-\gamma\cos(\omega t)]}
        \hat c_{m,n+1}^\dagger \hat c_{m,n}\\
        &+\mathrm{H.c.}
        \label{eq:FHH}
    \end{aligned}
\end{equation}
Under the Peierls substitution, the time-dependent hopping phase can be interpreted as a dimensionless vector potential
$A_y=2\pi\alpha m-\gamma\cos\omega t$. In the temporal gauge, this gives a uniform synthetic magnetic field
$B_z=2\pi\alpha$ and an ac synthetic electric field
$E_y=\gamma\omega\sin(\omega t)$. Therefore, the 1D Floquet AAH model can be mapped to an effective 2D HH model driven by an ac electric field. In the off-resonant regime, the high-frequency approximation of the effective Hamiltonian reads
\begin{equation}
    \begin{aligned}
        \hat H_{\mathrm{FHH}}^{(0)}
        ={}&
        -\sum_{m,n} J\,\hat c_{m+1,n}^\dagger \hat c_{m,n}\\
        &+\frac{\mathcal{J}_0(\gamma)\Delta}{2}\sum_{m,n}
        e^{i2\pi\alpha m}
        \hat c_{m,n+1}^\dagger \hat c_{m,n}\\
        &+\mathrm{H.c.}
        \label{eq:HH_eff}
    \end{aligned}
\end{equation}

\subsection{Landau-Zener tunneling}
    
The adiabatic theorem states that when the parameters are varied infinitely slowly and no level crossing occurs during the evolution, a system initially prepared in an instantaneous eigenstate remains in it. When the adiabatic condition is not strictly satisfied, the system may leak into other states during the evolution, with the transition probability described by the Landau-Zener formula~\cite{zener1932non,born1928beweis,wittig2005landau}
\begin{equation}
    P_{\mathrm{LZ}}
    =
    \exp\!\left(
    -\frac{\pi \Delta_{\rm gap}^2}{2 \left|d(\delta\epsilon)/dt\right|}
    \right),
    \label{eq:LZ}
\end{equation}
where the spectrum is locally simplified as two linearly crossing levels, $\left|d(\delta\epsilon)/dt\right|$ is the rate of change of the energy difference, and $\Delta_{\rm gap}$ is the minimum adiabatic gap.

In Thouless pumping, the energy difference near the crossing point varies with $\phi$, so that $\left|d(\delta\epsilon)/dt\right|$ can be decomposed as $\left|d(\delta\epsilon)/d\phi\right|\dot\phi$. For the AAH quasicrystal in the strongly localized regime, one may estimate $\left|d(\delta\epsilon)/d\phi\right|\sim \Delta$, $\dot\phi=2\pi/T\equiv\omega_0$, and $\Delta_{\rm gap}\sim J$. Therefore, in the strongly localized regime, the interbelt transition probability in the AAH quasicrystal is mainly determined by the temporal variation period of $\phi$, and can be approximated as
\begin{equation}
	P_{\mathrm{LZ}}
	\sim
	\exp\!\left(
	-\frac{J^2}{\omega_0\Delta  }
	\right).
	\label{eq:LZ_estimate}
\end{equation}
For $J^3/\Delta^2\ll\omega_0\ll J^2/\Delta$, the pumping is in a semi-adiabatic regime where it is adiabatic only with respect to the largest main gaps, whose magnitude is of order $J$, while the smaller gaps, suppressed by at least a factor on the order of $J/\Delta$, are crossed completely. This makes the dynamics of the strongly localized AAH model equivalent to an effective three-belt model. A longer pumping period can resolve small gaps, leading that more gaps become adiabatic and visible in the dynamics.
    
\subsection{Effective Floquet Hamiltonian in the off-resonant regime}

In a periodically driven system, the evolution at integer multiples of the driving period $T'$ is determined by the evolution operator over one period,
\begin{equation}
    \hat U_F = \mathcal{T} \exp\!\left[-i
    \int_0^{T'} \hat H(t)\, dt \right].
    \label{eq:UF}
\end{equation}
The spectrum of averaged dynamics is equivalently described by the effective Hamiltonian $\hat{H}_{\mathrm{eff}}$, defined through
$
\hat U_F = e^{-i \hat{H}_{\mathrm{eff}} T'}.
$

In the off-resonant regime with $\omega \gg \Delta$, the effective Hamiltonian can be expanded using the high-frequency expansion~\cite{goldman2014periodically}
\begin{align}
    &\hat{H}_{\mathrm{eff}}=\hat H_0
    +\sum_{n=1}^{\infty} \frac{1}{\omega n}
    \bigl[\hat V^{(n)}, \hat V^{(-n)}\bigr]
    \notag\\
    &\qquad
    +\sum_{n=1}^{\infty} \frac{\bigl[ [\hat V^{(n)}, \hat H_0], \hat V^{(-n)} \bigr]
        +
        \bigl[ [\hat V^{(-n)}, \hat H_0], \hat V^{(n)} \bigr]}{2\omega^2n^2}\notag\\
    &\qquad+ \mathcal{O}\!\left(\frac{1}{\omega^3}\right),
    \label{eq:Heff_expansion}
\end{align}
where
\begin{equation}
    \hat H(t) = \hat H_0 + \hat V(t), 
    \qquad
    \hat V(t) = 
    \sum_{n=1}^{\infty}
    \hat V^{(n)} e^{i n \omega t}
    +
    \hat V^{(-n)} e^{-i n \omega t}.
    \label{eq:fourier}
\end{equation}

For Thouless pumping, the slow linear phase $\omega_0 t$ can be treated as a constant within each driving period. The zeroth-order term of the high-frequency expansion then renormalizes the on-site modulation amplitude,
\begin{align}
    \hat{H}_0
    &= - J \sum_{j} \left( \hat{c}_{j+1}^\dagger \hat{c}_{j} + \text{H.c.} \right) \notag\\
    &\quad\, + \Delta \mathcal{J}_0(\gamma) \sum_{j} \cos\theta_j\, \hat{c}_{j}^\dagger \hat{c}_{j}
    \label{eq:Floquet_zeroth},
\end{align}
where $\theta_j=2\pi\alpha j - \omega_0 t$. The first-order term vanishes, and the second-order term induces a smaller correction to the tunneling amplitudes which is much smaller
\begin{align}
    \hat H_2
    ={}&
    4J\frac{\Delta^2}{\omega^2}\sin^2(\pi\alpha)
    \sum_j\left(
    \hat c_{j+1}^\dagger \hat c_j
    +
    \hat c_j^\dagger \hat c_{j+1}
    \right)\times
    \notag\\
    &
    \sum_{k=1}^{\infty}\Biggl[
    \frac{\mathcal{J}_{2k}^2(\gamma)}{(2k)^2}
    \sin^2(\theta_j+\pi\alpha)+\frac{\mathcal{J}_{2k-1}^2(\gamma)}{(2k-1)^2}
    \cos^2(\theta_j+\pi\alpha)
    \Biggr].
\end{align}
	
\section{\MakeUppercase{Numerical Methods and Parameters}}
\label{app:methods}
\subsection{Time evolution simulation}
We discretize the total evolution time into small intervals of size $\Delta t=T/N$ with total $N$ steps and approximate the evolution operator on each time by
\begin{equation}
    \ket{\psi(t+\Delta t)} \approx 
    \exp\!\left[-i\hat H(t)\Delta t\right]\ket{\psi(t)} .
\end{equation}
Since the tunneling matrix element is time independent, a Trotter decomposition is used to accelerate the time-evolution calculation, reducing the time complexity of each time step to $O(L)$, where $L$ is the system size.

The system size is chosen so that the wavepacket does not reach the boundaries during the entire simulation time. For the unmodulated quasicrystal, the time step is taken as $\Delta t \simeq T/10^4$, where $T$ is the pumping period. With Floquet modulation, $\Delta t \simeq T'/10^2 \simeq T/10^5$. We have checked that further reducing the time step does not produce obvious change in the dynamics, confirming the numerical convergence of our method.
    
\subsection{Calculation of the Floquet quasienergy spectrum}

Diagonalizing $U(T')$ gives the eigenvalues of $H_{\mathrm{eff}}$, namely the quasienergies $\epsilon_n$, and the corresponding Floquet eigenstates $\lvert n\rangle$.
\begin{equation}
    U(T') = \sum_n e^{-i \epsilon_n T'} \, \lvert n \rangle \langle n \rvert .
    \label{eq:evolution}
\end{equation}
Since the quasienergy is in the exponential, quasienergies differing by $\omega$ are equivalent. This equivalence of quasienergies originates from the discrete time-translation symmetry of the stroboscopic evolution, analogous to the first Brillouin zone in momentum space arising from the discrete translational symmetry of a lattice. A common gauge choice is $\epsilon_n \in [-\omega/2,\omega/2)$.

Eq.~\eqref{eq:UF} can be numerically calculated by discretizing the time evolution. Dividing one period into $N$ time steps with $\Delta t=T'/N$, 
\begin{equation}
    U(T') \approx \prod_{m=0}^{N-1} \exp\!\left[-i H(m\Delta t)\,\Delta t\right],
\end{equation}
where the product is ordered from right to left in increasing time. The quasienergy spectrum is also computed independently using the Sambe formalism~\cite{eckardt2017colloquium} in the extended Floquet Hilbert space. The quasienergy spectra obtained from the two methods are in good agreement.

\subsection{Floquet modulation and pumping parameters}
	
Throughout this work, energies are in units of the recoil energy of the short lattice $E_{\mathrm{r,s}}$, and time is in units of $\hbar/E_{\mathrm{r,s}}$. In typical experiments, the achievable depth of the short lattice is on the order of $V_{\mathrm{s}} \sim 10$, corresponding to a tunneling matrix element of order $J \sim 0.01$ in units of $E_{\mathrm{r,s}}$. Accordingly, unless otherwise specified, we fix the parameters to
\begin{equation}
    \Delta = V_{\mathrm{l}}/2 = 0.5, \qquad J = 0.02, \qquad V/J = 25 \gg 1.
\end{equation}

For the choice of pumping period, the semi-adiabatic regime requires
\begin{equation}
    J^2 T/\Delta \gg 1 .
    \label{eq:semi-adiabatic}
\end{equation}
We set $T = 15000$, corresponding to $J^2 T/\Delta = 12 \gg 1$.

For Fig.~\ref{fig:schematic}(d) in the main text, the lattice spacing ratio is chosen as $\alpha = \sqrt{2}/2$ without loss of generality.
The initial state uniformly occupies sites $35$ to $64$. The total system size is set to $100$ sites, ensuring that the wave packet does not reach the boundaries within $5T$.

For Fig.~\ref{fig:schematic}(c) in the main text, the spectrum is calculated using $1000$ lattice sites with $2000$ discretization steps. For Fig.~\ref{fig:velocity_alpha}(b) and (c), only the parameter $\alpha$ is varied, and the transport velocities are extracted by linear fitting of the pumping trajectory.

In the Floquet-modulated case Fig.~\ref{fig:diffusion} in the main text, the system size is set to $292$ sites, the initial state uniformly occupies $90$ sites and $T=34562$. In the semi-adiabatic regime, we choose $\omega = \omega_0/2$. In the resonant regime, $\omega$ is chosen to satisfy $J \ll \omega \lesssim \Delta$, and we use $\omega = 0.3$, which fulfills
$
\Delta \sim \omega,\ \omega \sim 15J.
$
In the off-resonant regime, the $s$-$p$ band gap is approximately estimated as $\Delta_{sp} \sim 2\sqrt{V E_{\mathrm{r,s}}} \approx 2\sqrt{10}$, and we take $\omega = 1.26$, such that
$\omega \sim 2.5V$ and $\Delta_{sp} \sim 5\omega$.

\section{\MakeUppercase{Numerical Results}}
\label{app:numerical}

\subsection{The relationship between pumping velocities and Chern numbers}
\label{app:population_velocity}

\begin{figure}[tbp]
    \centering
    \includegraphics[width=1\linewidth]{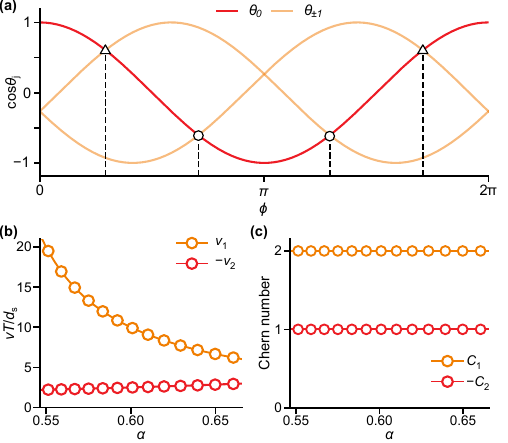}
    \caption{
        \textbf{Energy belt occupation and pumping velocities in optical quasicrystals}
        \textbf{(a)} 
        Tunneling between two neighboring sites happens when on-site energies equal. Here $\alpha=\sqrt{2}/2$, $\theta_{0}$, $\theta_{\pm1}$ are phases of the site (red line) and neighboring sites (orange line), and their intersections indicate the positions of gaps. These degeneracy points divide the interval $[0,2\pi)$ into three belts and the population fraction of each belt is given by the size of these angular sections.
        \textbf{(b)} Dependence of the pumping velocities of the two branches on $\alpha$ near $\alpha = \sqrt{2}/2$.
        \textbf{(c)} Chern numbers are calculated by the product of velocities and corresponding energy belt population fractions.
    }
    \label{fig:velocity_alpha}
\end{figure}

For $\Delta \gg J$, the spectrum can be understood by perturbation theory. Defining $\theta_j=(2\pi\alpha j-\phi)\bmod 2\pi$ for each site, the zeroth-order energy is $E_j^{(0)}=\Delta\cos\theta_j$. According to first-order perturbation theory, the hopping term couples two neighboring states with nearly degenerate energies
\begin{equation}
    \cos\theta = \cos(\theta + 2\pi\alpha).
    \label{eq:gap_eq}
\end{equation}
As a result, the two levels at $\Delta\cos\theta$ and $\Delta\cos(\theta+2\pi\alpha)$ split by an energy interval of $2J$~\cite{sinai1987anderson}. These degeneracies emerge in the spectrum as the two dominant gaps, which divide the spectrum into three belts. Since $\theta_j$ are uniformly distributed over $[0,2\pi)$ for irrational $\alpha$~\cite{weyl1916gleichverteilung}, the populations of different belts are calculated from the angle between the gap-opening positions in Fig.~\ref{fig:velocity_alpha}(a). For $\alpha > 1/2$, the population ratio is
\begin{equation}
    1-\alpha : 2\alpha-1 : 1-\alpha.
    \label{eq:population}
\end{equation}

The pumped charge is given by the Chern number. The corresponding current is proportional to the pumping velocity and the belt population. Under uniform occupation, the populations of the belts with Chern numbers $C_1=+2$ and $C_2=-1$ are given by Eq.~\eqref{eq:population}. Their corresponding velocities are therefore
\begin{equation}
    v_1=\frac{2}{2\alpha-1}, \qquad
    v_2=-\frac{1}{1-\alpha}.
    \label{eq:velocity}
\end{equation}
Fig.~\ref{fig:velocity_alpha}(b) shows that the velocities (open circles) extracted by fitting the evolution for different $\alpha$ agree very well with the theoretical curves (solid lines). The Chern numbers extracted inversely from these velocities are consistent with the gap labeling theorem in Fig.~\ref{fig:velocity_alpha}(c). Notably, as $\alpha$ approaches $1/2$, the population of the belt with $C=2$ tends to zero, causing the corresponding pumping velocity to diverge.

From a dynamical perspective, atoms are almost completely reflected at the edges of the three belts, while they tunnel through the smaller gaps. Therefore, whenever the system passes through a degeneracy, the atom hops to the adjacent site, with the direction dictated by staying on the same adiabatic belt. From this picture, we can calculate the pumping velocity of each belt. The hopping events occur after equal phase intervals, which also explains the constant velocity.

\subsection{Unfolding Floquet quasi-energy spectrum}
\label{app:gauge}

In a Floquet system, the evolution at integer multiples of the driving period has a discrete time periodicity. Quasienergies differing by $\omega$ generate the same stroboscopic evolution. In the off-resonant regime, the first Floquet zone is sufficiently wide, so spectra from different quasienergy zones do not overlap in Fig.~\ref{fig:Quasi_spectrum}(a). The quasienergy spectrum is then well described by the zeroth-order approximation in Eq.~\eqref{eq:Floquet_zeroth}. In the resonant regime, however, quasienergy spectra from different Floquet zones overlap.

\begin{figure}[tbp]
\centering
\includegraphics[width=1\linewidth]{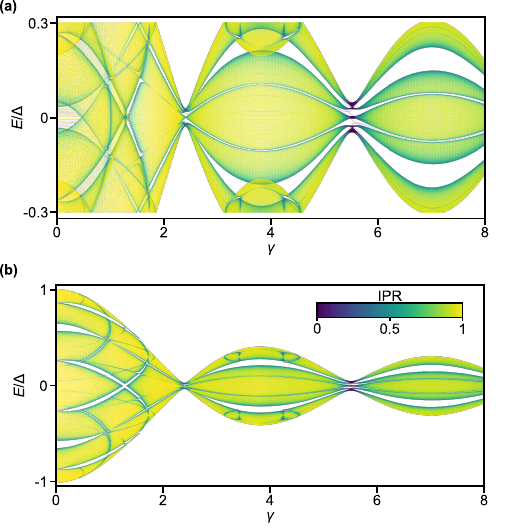}
\caption{\textbf{Folded and unfolded quasienergy spectrum at $\omega=0.3$.}
\textbf{(a)} Folded quasienergy spectrum in the first Floquet zone $[-\omega/2,\omega/2)$, in which quasienergy from different zones overlap.
\textbf{(b)} Unfolded quasienergy spectrum by choosing the extended Floquet-zone gauge. The color indicates IPR.
}
\label{fig:unfold}
\end{figure}

Here we employ an extended Floquet-zone gauge to unfold the quasienergy spectrum from the first Floquet zone, in analogy with unfolding bands in an extended Brillouin zone. 
The detail of unfolding is by choosing an integer $m_n$ such that the shifted quasienergy $\epsilon_n+m_n\omega$ is closest to the target energy $E_{j_n}$ associated with this site 
\begin{equation}
    m_n=
    \arg\min_m
    \left|
    \epsilon_n+m\omega
    -
    E_{j_n}
    \right|.
\end{equation}
When the target energy is taken from the zeroth-order high-frequency approximation $E_{j_n}=\mathcal{J}_0(\gamma)\Delta\cos\theta_{j_n}$, the unfolded quasienergy spectrum in the resonant regime shows a clear structure in Fig.~\ref{fig:unfold}. 
Floquet driving can open additional gaps that close and reopen as $\gamma$ varies, leading to the change of the topology of the quasienergy spectrum.

We first use the inverse participation ratio (IPR) to cross check the unfolding. The resonant regime can be regarded as including higher-order driving-induced perturbative corrections beyond the off-resonant regime. Quasienergies with eigenstates similar to those in the off-resonant case should remain close to each other after unfolding. Each quasienergy eigenstate $|n\rangle$ can be associated with a site position through its largest overlap with the local basis states $|j\rangle$,
\begin{equation}
    j_n=\arg\max_j |\langle j|n\rangle|^2 .
\end{equation}
This definition is valid when the system remains strongly localized in the resonant regime. For a normalized quasienergy eigenstate $|n\rangle$, the IPR is defined as
\begin{equation}
    \mathrm{IPR}_n=\sum_j |\langle j|n\rangle|^4 .
    \label{eq:IPR}
\end{equation}
Fig.~\ref{fig:unfold} also shows that the IPR is close to 1 in most regions, indicating that the system remains localized and justifying the the extended Floquet-zone gauge to unfold the energy spectrum.

\subsection{Gap opening in the Floquet quasienergy spectrum}
\label{app:gap}

\begin{figure}[t]
\centering
\includegraphics[width=1\linewidth]{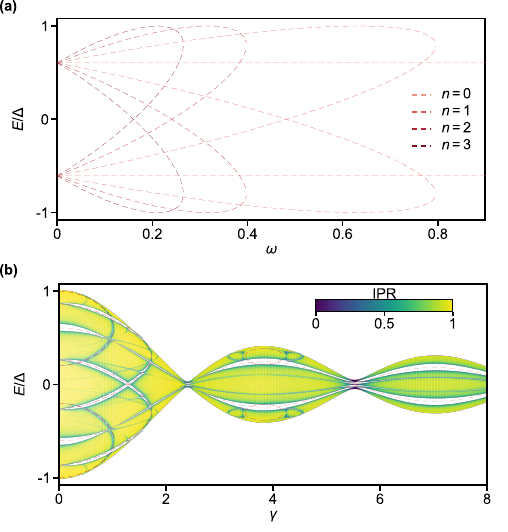}
\caption{
    \textbf{Positions of gap openings for different orders of Floquet coupling in the resonant regime at $\omega=0.3$.}
    \textbf{(a)} Gap-opening positions in the $\gamma \to 0$ limit for different orders of Floquet coupling. For the unmodulated case $n=0$, the positions are independent of $\omega$. Higher-order couplings have smaller cutoff frequencies and exist only in the lower-$\omega$ region.
    \textbf{(b)} Gaps open in the resonant regime at $\omega=0.3$. The solutions with $n=0,\pm1,\pm2$ exist, and gaps open only in the region where $|\mathcal{J}_0(\gamma)|$ is sufficiently large. The predicted opening agrees well with gaps in the calculated unfolded quasienergy spectrum.
}
\label{fig:gap_opening}
\end{figure}

In the absence of modulation, the gap-opening positions are determined by Eq.~\eqref{eq:gap_eq}, which corresponds to the condition that two neighboring sites have equal onsite energies. Once Floquet modulation is introduced, neighboring sites whose energy difference equals an integer multiple of $\omega$ can also be coupled. This leads to the extended gap equation
\begin{equation}
    \cos\theta-\cos(\theta+2\pi\alpha)=\frac{n\omega}{\Delta},
    \label{eq:extended_gap_eq_gamma0}
\end{equation}
where $n$ denotes the order of the Floquet coupling and $\Delta\cos\theta$ denotes the on-site energy. The case $n=0$ is the unmodulated quasicrystal. For positive and negative values of $n$, the two corresponding pairs of solutions give the gap-opening positions of different Floquet-coupling orders in the limit $\gamma\to0$, and the spacing between each pair of solutions is $|n|\omega$. Since the absolute value of the left-hand side of Eq.~\eqref{eq:extended_gap_eq_gamma0} has a maximum value of $2\sin(\pi\alpha)$, except for the case $n=0$, each coupling order has a cutoff frequency as shown in Fig.~\ref{fig:gap_opening}(a). Above this cutoff frequency, the corresponding Floquet coupling cannot open a gap. The cutoff decreases for higher orders, since it becomes increasingly difficult to find two spectral points within a beltwidth of order $\Delta$ whose energy difference equals $n\omega$.
Under the Floquet modulation and high frequency approximation, the Hamiltonian can be approximated as the zeroth-order high-frequency Floquet Hamiltonian $\hat{H}_0$. The corresponding gap equation, considering both the modulation frequency and amplitude, becomes
\begin{equation}
    \cos\theta-\cos(\theta+2\pi\alpha)
    = \frac{n\omega}{\mathcal{J}_0(\gamma)\Delta}.
    \label{eq:extended_gap_eq}
\end{equation}
For a fixed resonant frequency $\omega=0.3$, the gap lines predicted by this equation are plotted on the quasienergy spectrum in Fig.~\ref{fig:gap_opening}(b). 
Missing gaps are due to the relevant states removed by bare quasicrystal gaps and 
higher order coupling effects.

Moreover, Eq.~\eqref{eq:extended_gap_eq} gives the cutoff condition for the $n$th-order Floquet coupling in the two-parameter space of $\omega$ and $\gamma$
\begin{equation}
    \frac{\omega}{\Delta}
    = \frac{2\sin(\pi\alpha)}{|n|}\,|\mathcal{J}_0(\gamma)|,
    \label{eq:cut_off}
\end{equation}
which coincides with the $|\mathcal{J}_0|$-shaped envelope observed in the average IPR. This cutoff line gives the boundary for the opening of the gaps associated with the $n$th-order Floquet coupling, above which the corresponding gaps are absent and the associated fast-pumping channels disappear.

Finally, the widths of the gaps are determined by the coupling strengths. Different order Floquet coupling strengths can be obtained from the Fourier expansion in Eq.~\eqref{eq:fourier}, with components
\begin{equation}
    \begin{aligned}
        \hat{V}^{(2k)}
        &= (-1)^k\mathcal{J}_{2k}(\gamma)\, \Delta \sum_j \cos\theta_j \, c_j^\dagger c_j, \\
        \hat{V}^{(2k+1)}
        &= (-1)^k\,\mathcal{J}_{2k+1}(\gamma)\, \Delta \sum_j \sin\theta_j \, c_j^\dagger c_j,
    \end{aligned}
    \label{eq:fourier_components}
\end{equation}
where $\theta_j=2\pi\alpha j-\omega_0t$. The $n$th-order coupling Hamiltonian is therefore proportional to $\mathcal{J}_n(\gamma)$. This explains why higher-order couplings are weaker and open smaller gaps, and also why gaps of different orders generally exhibit opening and closing points as $\gamma$ varies, corresponding to the zeros of the Bessel functions of different orders.

\subsection{Dynamics in different frequency regimes}
\label{app:dynamics}

In the semi-adiabatic regime with small $\gamma$, the Landau-Zener tunneling condition for $H(t)$ remains approximately valid as long as the modulation-induced correction to the sweep rate is small, $\dot{\phi}\simeq\omega_0+\gamma\omega\simeq\omega_0$. The periodic modulation of $\phi(t)$ mainly shifts the tunneling positions along the pumping trajectory. 
	
\begin{figure*}[ht]
    \centering
    \includegraphics[width=\linewidth]{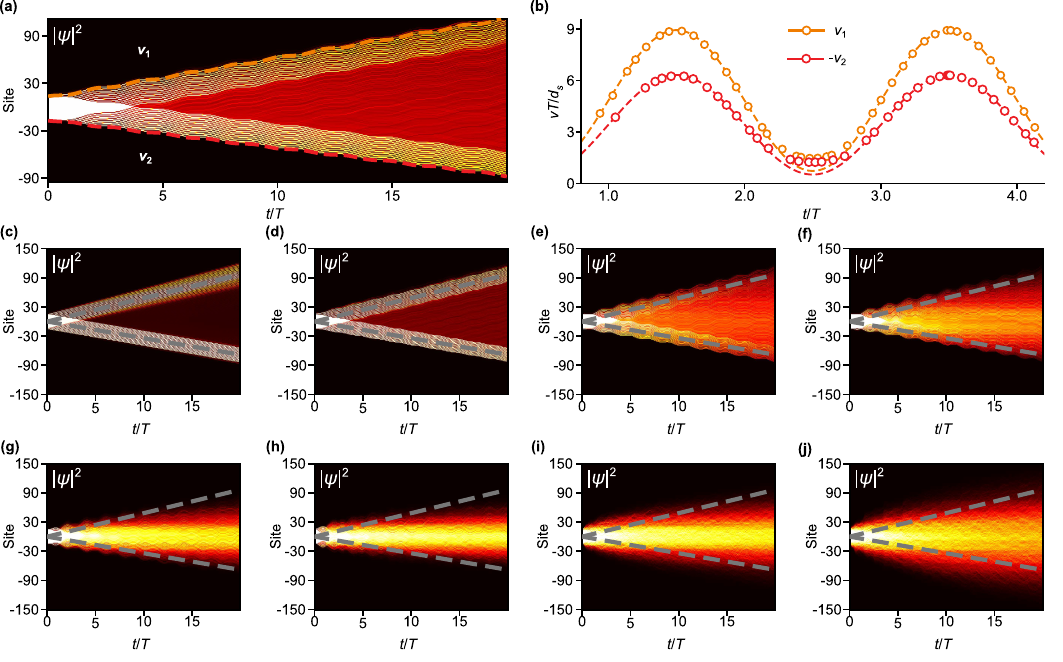}
    \caption{
        \textbf{Dynamics in the semi-adiabatic regime at $\omega = 0.5\omega_0$.}
        \textbf{(a)} Oscillatory dynamics of the semi-adiabatic pumping at $\gamma=1.7$. Red and orange dashed curves are the boundaries extracted from pumping trajectories.
        \textbf{(b)} Velocities obtained by statistically extracting the intervals between tunneling events, together with the theoretical sinusoidal curves derived from $\dot{\phi}$. The velocities oscillate around their original values with amplitudes proportional to $\omega$ and $\gamma$. The small dependency near the valleys arises from the slow-velocity regions requiring even denser sampling in spectral space. 
        \textbf{(c)--(j)} are $\gamma = 0,\,1,\,3,\,4,\,6,\,8,\,30,\,70$. The gray dashed lines indicate pumping at $v_1$ and $v_2$ of unmodulated quasicrystal.
    }
    \label{fig:low_frequency}
\end{figure*}

When the phase is no longer purely linear and a semi-adiabatic modulation is imposed, the tunneling events shift from their previously uniform temporal spacing and the variation occurs at the modulation frequency $\omega$. Since the hopping interval is inversely proportional to the pumping velocity, the velocity of each belt also oscillates at the same frequency $\omega$. Taking the unmodulated velocity $v_0$ as the reference, the instantaneous velocity can be expressed as $v(t)=v_0\left(1-\gamma \frac{\omega}{\omega_0}\sin(\omega t)\right)$. Fig.~\ref{fig:low_frequency}(b) shows oscillations in the pumping velocity for two belts with a period $2T$, accompanied by an envelope modulation of the trajectory shown in Fig.~\ref{fig:low_frequency}(a), corresponding to $\omega=\omega_0/2$. The deviation between the pumping velocity extracted from Fig.~\ref{fig:low_frequency}(a) and the theoretical curve near the valleys originates from the extremely slow dynamics in this regime. When pumping intervals become very long, an infinitely dense sampling of initial positions would be required to obtain a reliable averaged velocity. Noticeable discrepancies in the valley regions arise from the finite number of initially occupied sites in the simulation.

With increasing $\gamma$, the semi-adiabatic condition gradually breaks down. Because the instantaneous pumping velocity becomes too large, particles undergo interbelt transitions near the degeneracy points, thereby destroying the original stable pumping trajectories. In the limit of very large $\gamma$, such tunneling events can be regarded as approximately random with respect to the initial state distribution, leading to a broad spreading of the density over the region bounded by the gray dashed envelope lines.

\begin{figure*}[ht]
\centering
\includegraphics[width=\linewidth]{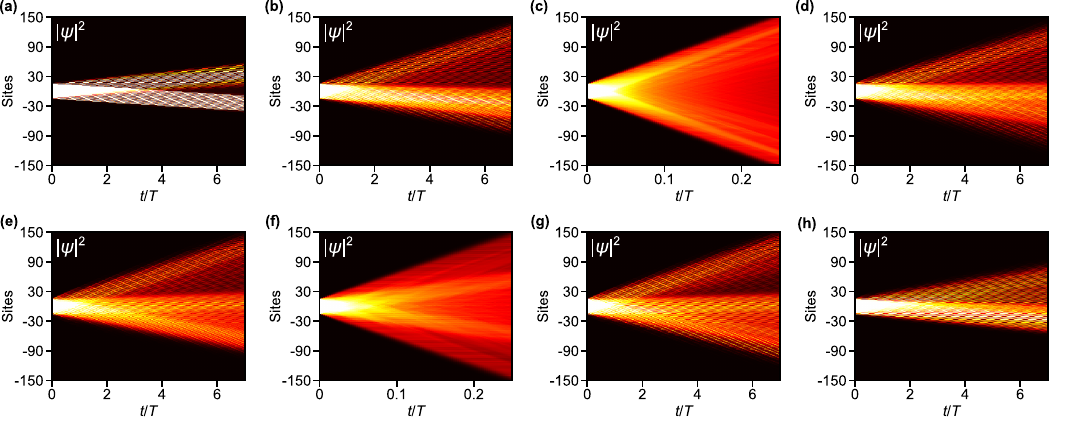}
\caption{
    \textbf{Dynamics in the off-resonant regime at $\omega \approx 1.26$.}
    \textbf{(a)--(h)} are $\gamma = 1,\,2,\,2.4,\,2.8,\,5,\,5.6,\,6.1,\,7.2$. Here pumping at $\gamma=1$ is similar to the unmodulated quasicrystal. $\gamma=2.4$ and $\gamma=5.6$ are near the zeros of $\mathcal{J}_0$, and the corresponding pumping dynamics are strongly diffusive. Others close to the zeros remain localized, in which the $G=\pm2$ gaps become more pronounced and higher velocities emerge.
}
\label{fig:high_frequency}
\end{figure*}

In the off-resonant regime near the zeros of $\mathcal{J}_0$, the onsite potential in $H_0$ is effectively suppressed, so that the system delocalizes and exhibits strong diffusion in Fig.~\ref{fig:high_frequency}(c) and (f). At other values of $\gamma$, in Fig.~\ref{fig:Quasi_spectrum}(a), the system remains close to the zeros of $\mathcal{J}_0$ but has not yet fully left the localized phase. In this regime, the additional gaps with labels $\pm 2$ gradually become more pronounced, and the velocities associated with the corresponding Chern numbers also emerge more clearly in the other panels.

\clearpage
\subsection{IPR extracted from Floquet quasienergy eigenstates}

\begin{figure}[h]
\centering
\includegraphics[width=\linewidth]{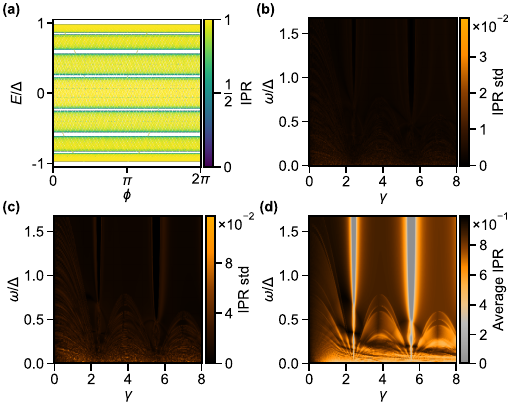}
\caption{
    \textbf{IPR robustly reflects the topological energy spectrum in Floquet quasicrystals.}
    \textbf{(a)} Quasienergy spectrum for different $\phi$ at $\omega=0.3$ and $\gamma=0.4$, similar to the flat energy spectrum of the unmodulated quasicrystal. The color indicates the IPR and also remains almost independent of $\phi$.
    \textbf{(b)} IPR standard deviation over different $\phi$ is small, at the level of $10^{-2}$.
    \textbf{(c)} IPR standard deviation for the initial state uniformly occupying 30 sites. The standard deviation is calculated over different $\phi$ of the average IPR, which is calculated based on the weights of Floquet quasienergy eigenstates.
    \textbf{(d)} The average IPR for the initial state at $\phi=0$ agrees well with Fig.~\ref{fig:diffusion}(c) in the main text.
}
\label{fig:IPR_std}
\end{figure}

Averaging the IPR over all quasienergy eigenstates at each $(\omega,\gamma)$ point for $\phi=0$ gives Fig.~\ref{fig:diffusion}(c). To verify that the average IPR obtained from the $\phi=0$ spectrum can represent the average over all $\phi$, we further calculate the quasienergy spectrum and the dependence of the corresponding IPR on $\phi$. Similar to the flat-belt property of the unmodulated quasicrystal spectrum resulting from unitary periodicity, the modulated quasienergy spectrum also remains flat along $\phi$ in Fig.~\ref{fig:IPR_std}(a). The standard deviation of the IPR over $\phi$ is also small in Fig.~\ref{fig:IPR_std}(b).
	
For the initial state uniformly occupying 30 sites, we further calculate the average IPR and its standard deviation from the decomposition of the initial state in the Floquet quasienergy eigenbasis. This confirms that the initial state is also approximately uniformly distributed over the Floquet quasienergy eigenstates. As a result, the averaged IPR obtained for the initial state in Fig.~\ref{fig:IPR_std}(d) agrees well with the average over all eigenstates in Fig.~\ref{fig:diffusion}(c).

\begin{figure}[ht]
\centering
\includegraphics[width=\linewidth]{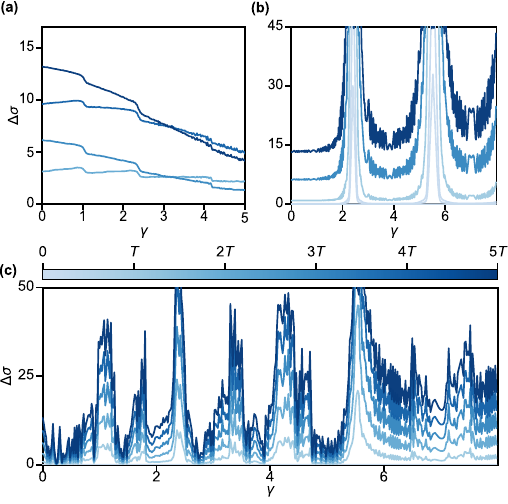}
\caption{
    \textbf{Representative width growth $\Delta\sigma(t)$ under different modulation strength $\gamma$ in different regimes.} 
    \textbf{(a)} The semi-adiabatic regime, i.e. low-frequency modulation regime. The four curves correspond to $2T$, $3T$, $4T$, and $5T$. Curves separated by $2T$ show similarity, in agreement with the modulation frequency $\omega=\omega_0/2$.
    \textbf{(b)} The off-resonant regime i.e. high-frequency modulation regime at $\omega\approx1.26$. Fast expansion happens near zeros of $\mathcal{J}_0(\gamma)$, whereas for the remaining regimes, $\Delta\sigma(t)$ increase linearly, similar to the unmodulated quasienergy.
    \textbf{(c)} The resonant regime at $\omega\approx0.25$, in which multiple peaks away from the zeros of $\mathcal{J}_0(\gamma)$ exhibit rapid width growth and reflect fast pumping induced by Floquet couplings.
}
\label{fig:Delta_sigma}
\end{figure}

\makeatletter
\def\bibsection{%
  \par
  \begingroup
    \baselineskip26\p@
    \bib@device{\hsize}{72\p@}%
  \endgroup
  \nobreak\@nobreaktrue
  \addvspace{19\p@}%
}
\makeatother

\bibliography{QP.bib}

\end{document}